\documentclass[11pt]{article}
\usepackage{acl}
\usepackage{booktabs}
\usepackage{tikz}
\usetikzlibrary{arrows.meta,positioning,calc,fit,shapes.geometric,decorations.pathreplacing,backgrounds}
\usepackage{xcolor}
\usepackage{times}
\usepackage{latexsym}
\usepackage{amsmath}
\usepackage{bm}
\usepackage{braket}
\usepackage{multirow} 
\usepackage{mathtools}
\usepackage{amssymb}
\usepackage{tcolorbox}

\usepackage{fancyhdr}
\usepackage{afterpage}

\fancypagestyle{firstpage}{
  \fancyhf{} 
  \fancyfoot[L]{\scriptsize\textsuperscript{*}Equal contribution \quad \textsuperscript{†}Corresponding author}
}

\tcbuselibrary{breakable, skins, theorems}

\usepackage[T1]{fontenc}

\usepackage[utf8]{inputenc}

\usepackage{microtype}

\usepackage{inconsolata}

\usepackage{graphicx}

\title{CLAQS: Compact Learnable All‑Quantum Token Mixer with Shared‑ansatz for Text Classification}

\author{
 \textbf{Junhao Chen\textsuperscript{1,*}},
 \textbf{Yifan Zhou\textsuperscript{1,*}},
 \textbf{Hanqi Jiang\textsuperscript{1}},
 \textbf{Yi Pan\textsuperscript{1}},
\\
 \textbf{Yiwei Li\textsuperscript{1}},
 \textbf{Huaqin Zhao\textsuperscript{1}},
 \textbf{Wei Zhang\textsuperscript{2}},
 \textbf{Yingfeng Wang\textsuperscript{3, †}},
 \textbf{Tianming Liu \textsuperscript{1,†}}
\\
 \textsuperscript{1}University of Georgia,
 \textsuperscript{2}Augusta University,
 \textsuperscript{3}University of Tennessee at Chattanooga\\
\\
}

\begin{document}
\maketitle
\begin{abstract}
Quantum compute is scaling fast, from cloud QPUs to high throughput GPU simulators, making it timely to prototype quantum NLP beyond toy tasks. However, devices remain qubit limited and depth limited, training can be unstable, and classical attention is compute and memory heavy. This motivates compact, phase aware quantum token mixers that stabilize amplitudes and scale to long sequences. We present CLAQS, a compact, fully quantum token mixer for text classification that jointly learns complex-valued mixing and nonlinear transformations within a unified quantum circuit. To enable stable end-to-end optimization, we apply $\ell^1$ normalization to regulate amplitude scaling and introduce a two-stage parameterized quantum architecture that decouples shared token embeddings from a window-level quantum feed-forward module. Operating under a sliding-window regime with document-level aggregation, CLAQS requires only eight data qubits and shallow circuits, yet achieves 91.64\% accuracy on SST-2 and 87.08\% on IMDB, outperforming both classical Transformer baselines and strong hybrid quantum–classical counterparts. 
\end{abstract}

\section{Introduction}
\begin{figure*}[t]
    \centering
    \includegraphics[width=\textwidth]{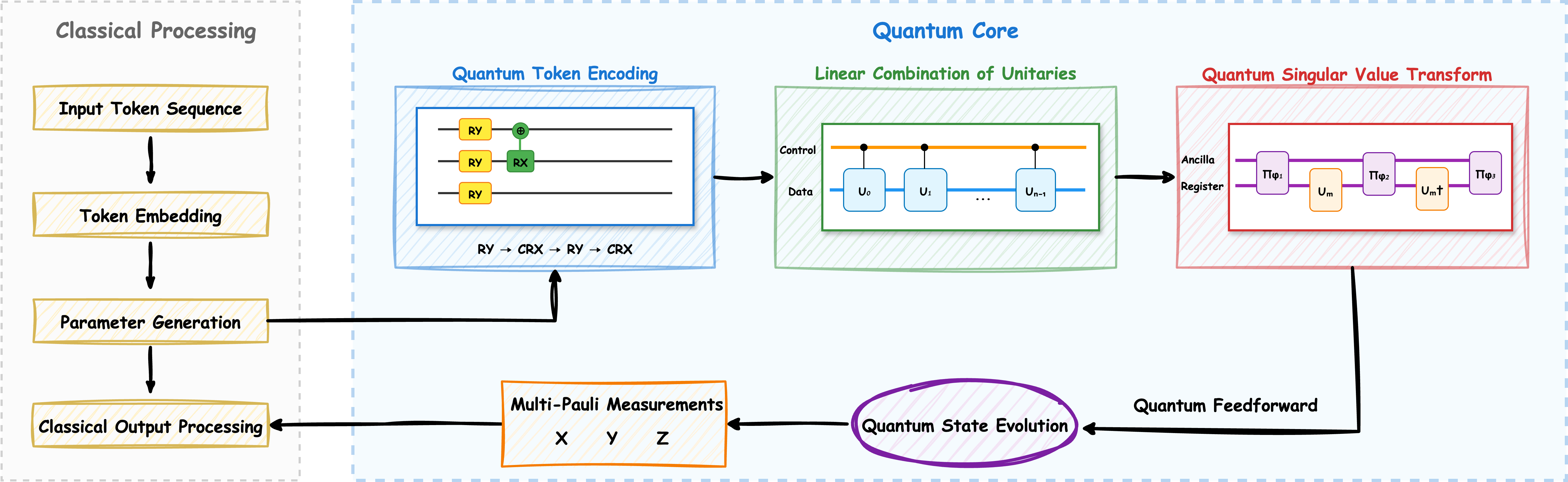}
    \caption{\textbf{Overall architecture of the CLAQS model.} 
    The model integrates token-shared quantum embedding, LCU–QSVT mixing, and a quantum feed-forward block, 
    mirroring the structure of a Transformer layer while operating on quantum representations.}
    \label{fig:architecture}
\end{figure*}

Transformer-based language models have revolutionized natural language processing (NLP) by introducing the self-attention mechanism, which enables flexible token interactions and drives state-of-the-art performance across a wide range of tasks~\citep{vaswani2017attention}. 
However, attention-based architectures come at the cost of massive parameter counts and quadratic computational complexity, motivating the search for more efficient sequence mixers. 
Recent studies demonstrate that high-quality representations can arise from alternative mixing operations, such as Fourier or convolutional transforms~\citep{lee-thorp-etal-2022-fnet}, suggesting that attention may not be the only viable paradigm for effective token integration.

Meanwhile, advances in quantum computing have opened new directions for sequence modeling. 
Quantum systems can represent information in exponentially large Hilbert spaces via superposition and entanglement, offering a unique mechanism for encoding complex linguistic dependencies beyond classical vector spaces. 
Yet, Noisy Intermediate-Scale Quantum (NISQ) hardware restricts circuit depth and qubit count, making fully quantum language models infeasible at scale. 
This has spurred the development of hybrid quantum–classical approaches, such as Adaptive Quantum–Classical Fusion (AQCF)~\citep{pan2025bridging}, which dynamically alternate between quantum and classical processing. 
However, these methods often rely on static quantum components and do not fully exploit the potential of learnable quantum representations.

In this paper, we propose \textbf{CLAQS}, a compact, end-to-end quantum–classical model designed for text classification under realistic NISQ constraints. Our key innovations are threefold:
\begin{itemize}
    \item \textbf{Learnable quantum token mixing:} CLAQS parameterizes both the quantum mixing weights and nonlinear transformations as complex-valued parameters, enabling data-driven discovery of optimal token interactions instead of relying on analytically fixed coefficients.
    \item \textbf{Stabilized training via normalization:} An $\ell_1$ normalization constraint on mixing amplitudes ensures stable end-to-end optimization and mitigates coefficient explosion or vanishing, which were major limitations in earlier quantum designs.
    \item \textbf{Dual-stage quantum circuit architecture:} A shared token-level embedding ansatz and a window-level quantum feed-forward block jointly decouple representation construction and consumption—mirroring the structure of Transformer layers while maintaining qubit efficiency.
\end{itemize}

With only eight data qubits and shallow circuits, CLAQS achieves \textbf{91.64\%} accuracy on SST-2 and \textbf{87.08\%} on IMDB, outperforming both classical Transformers and strong hybrid baselines.
To our knowledge, this is the first demonstration that a compact, learnable quantum circuit can act as an effective and stable sequence mixer for NLP, bridging the gap between classical attention and quantum superposition under tight resource budgets.

\section{Related Work}
\paragraph{Classical attention and alternative token mixers.}
The Transformer popularized self-attention and became the dominant sequence mixer in NLP~\citep{vaswani2017attention}. To reduce attention’s quadratic cost, FNet mixes tokens with fixed Fourier transforms while remaining competitive on language tasks~\citep{lee-thorp-etal-2022-fnet}. Beyond Fourier, recent \emph{subquadratic} mixers—long-convolution operators and selective state-space models—have closed much of the quality gap to attention while improving scaling; representative examples include Hyena (implicit long convolutions with data-gated kernels) and Mamba (selective state spaces with input-dependent dynamics)~\citep{poli2023hyena,gu2023mamba}. These efficiency trends motivate re-examining token mixing on alternative computational substrates, including quantum circuits.

\paragraph{Non-attention quantum text classifiers and encoders.}
Parallel to attention-based routes, a separate line studies end-to-end text classification without Transformer-style attention. Quantum-like semantic models have been proposed for sentiment and topic classification~\citep{gruzdeva2025quantum}, while quantum n-gram language models target tweet categorization~\citep{payares2023quantum}. On the representation side, task-oriented quantum text encoding, using amplitude-encoded features with QSVM back-ends, offers alternative front-ends for small-scale classification~\citep{alexander2022quantum}. These efforts contextualize our focus on a fully learnable LCU–QSVT token mixer that can be trained end-to-end and potentially plugged into larger classical pipelines.

\paragraph{Fully quantum NLP and quantum modules in classical models.}
Compositional QNLP (DisCoCirc) provides a circuit-level account of meaning~\citep{coecke2021mathematics}, with hardware demonstrations such as grammar-aware sentence classification and executions of compositional models on NISQ devices~\citep{meichanetzidis2023grammar,lorenz2023qnlp}. Closer to Transformers, quantum modules have been \emph{inserted} into attention: QKSAN computes kernelized similarities on quantum circuits within self-attention~\citep{zhao2024qksan}, and hybrid/quantized attention has been explored for molecular sequences~\citep{smaldone2025hybrid}. At the system level, dynamic hybrid co-design allocates quantum resources by input complexity~\citep{pan2025bridging}, while \citet{yang2022bert} attach quantum temporal convolutions to BERT—illustrating a pragmatic near-term path where small quantum sub-networks complement classical models.

\paragraph{Quantum-native attention mechanisms.}
Beyond replacing classical attention, several works implement token mixing directly in Hilbert space. QSANN introduces Gaussian-projected quantum self-attention for text classification~\citep{li2024quantum}. Quantum Vision Transformers define fully quantum attention layers (including compound-matrix attention) and argue asymptotic advantages in runtime and parameter count~\citep{cherrat2024quantum}. QMSAN computes query–key similarities between mixed quantum states via SWAP-test-style overlaps~\citep{chen2025quantum}. In this vein, \emph{Quixer} prepares an LCU superposition over token states and applies a QSVT polynomial as a learned nonlinearity—effectively realizing attention in a quantum register~\citep{Khatri2024Quixer}. Our \textbf{CLAQS} follows this quantum-native mixer direction for supervised classification and contributes a compact, fully learnable circuit: (i) both LCU mixing weights and the QSVT polynomial are trained end-to-end, (ii) LCU coefficients are $\ell_1$-normalized to stabilize training, and (iii) a shared token-embedding ansatz is decoupled from a window-level quantum feed-forward module, yielding strong accuracy under an 8-data-qubit budget. \footnote{In classical simulation we evaluate $P_c(M)$ by repeated applications of $M$ rather than materializing full block encodings, following Quixer’s classical-simulation strategy and our implementation notes.}

\section{Method}
\subsection{Problem Setup and Design Goals}
We study supervised text classification with input token sequence $x=(w_0,\ldots,w_{n-1})$ and label $y\in\{1,\ldots,C\}$. Our goal is to replace dot-product attention with a \emph{compact quantum mixer} that is trainable end-to-end, uses only $q{=}8$ data qubits, and preserves the principled \emph{LCU} $\rightarrow$ \emph{QSVT} $\rightarrow$ feed-forward $\rightarrow$ measurement pipeline introduced in Quixer while adapting it to classification tasks. Our design goals are (G1) compactness under strict qubit/depth budgets, (G2) phase-aware token interactions with explicit nonlinearity, (G3) stable optimization with regulated amplitudes, and (G4) scalability to long documents via sliding-window aggregation.

\begin{figure*}[t]
  \centering
  \resizebox{0.9\linewidth}{!}{%
    \includegraphics{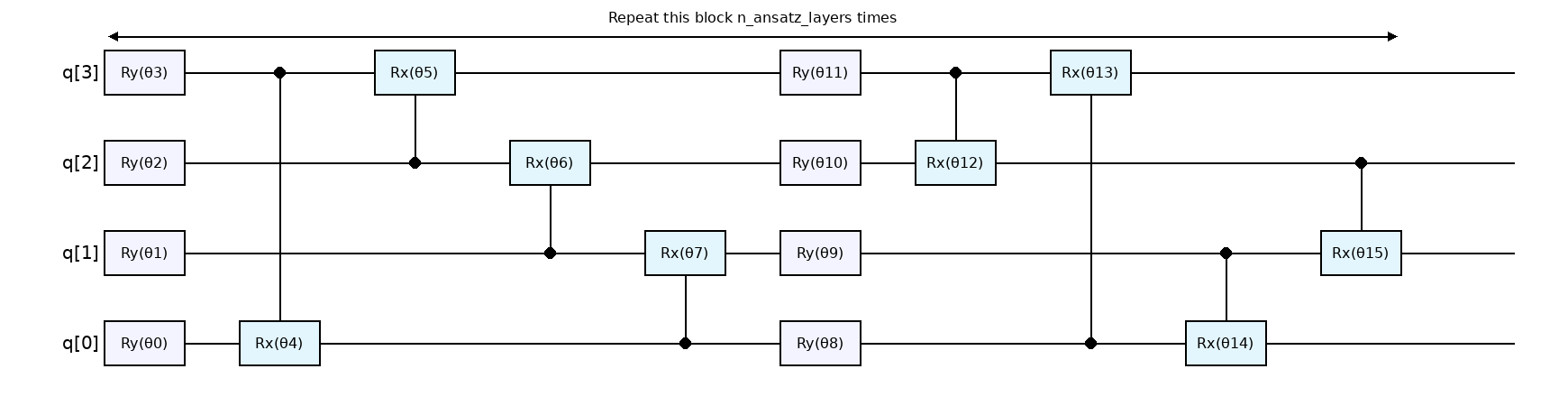}
  }
  \caption{The detailed one layer of the shared \textit{ansatz-14} for quantum token encoding (repeated $\ell$ times).}
  \label{fig:ansatz14}
\end{figure*}
\subsection{Model Overview}
While maintaining the above pipeline, CLAQS re-parameterizes all components to be trainable and differentiable, yielding a fully end-to-end learnable quantum circuit. The model proceeds in five stages. 
First, it maps each token embedding into a unitary circuit through a shared token-level ansatz. 
Second, it performs a complex-weighted unitary superposition with learnable parameters, normalized to ensure stability. 
Third, a trainable polynomial transformation acts as a quantum nonlinearity. 
Fourth, an independent feed-forward PQC refines the aggregated state at the window level. 
Finally, multi-axis measurements produce classical features for downstream classification through a lightweight MLP head.

\subsection{Unitary Token Embedding}
Let $\mathbf e_w\in\mathbb{R}^{d_e}$ denote a classical token embedding. We map it to angles $\boldsymbol\theta_w = W_E\,\mathbf e_w$ (with Xavier initialization for $W_E$), which parameterize a $q$‑qubit \textit{ansatz‑14} PQC $U(\boldsymbol\theta_w)$. As illustrated in Figure~\ref{fig:ansatz14}, a single \textit{ansatz‑14} layer alternates $RY$ and controlled‑$RX$ gates and is repeated $\ell$ times; with $q$ qubits this yields $4\ell q$ trainable angles while keeping depth modest. The circuit template is \emph{shared} across tokens, where each token has its own $\boldsymbol\theta_w$ but the same gate pattern, improving sample efficiency and aligning with our unitary token‑embedding design.

\subsection{LCU Mixer with Learnable Complex Coefficients and L1 Normalization}
For the window $\{U_j\}_{j=0}^{n-1}$ of token unitaries, we construct a linear combination of unitaries
\begin{equation}
M(\boldsymbol b)=\sum_{j=0}^{n-1} b_j\, U_j,\qquad \boldsymbol b\in\mathbb{C}^n,
\label{eq:lcu}
\end{equation}
where $\boldsymbol b$ is trainable and complex, representing the quantum magnitude and phase. To stabilize training and respect block-encoding assumptions, we apply L$^1$ normalization per forward pass:
\begin{equation}
\tilde b_j=\frac{b_j}{\sum_{k=0}^{n-1} |b_k|}\quad\Rightarrow\quad \sum_{j=0}^{n-1} |\tilde b_j|=1.
\label{eq:l1}
\end{equation}
On hardware, $M(\boldsymbol b)$ is realized using a selection unitary $U_{\text{SEL}}$ and a preparation unitary $U_{\text{PREP}}$ with post-selection on the control register; in classical simulation we compute \eqref{eq:lcu} directly, following Quixer’s implementation strategy.%

\subsection{QSVT Nonlinearity with Learnable Polynomial}
We apply Quantum Singular Value Transformation (QSVT) to the block-encoded $M(\tilde{\boldsymbol b})$, effecting a polynomial transform
\begin{equation}
P_{\boldsymbol c}(M)=\sum_{k=0}^{d} c_k\, M^k,
\label{eq:qsvt}
\end{equation}
with learnable degree $d$ and coefficients $\boldsymbol c$. Operationally, QSVT is implemented via a sequence of controlled phase rotations interleaved with $U_M$ and $U_M^\dagger$; in classical simulation we evaluate the polynomial by repeated applications of $M$. Making $\boldsymbol c$ learnable allows the model to tailor its nonlinearity to data, rather than fixing $P_{\boldsymbol c}$ \emph{a priori}. This follows the Quixer design where a QSVT polynomial acts as a nonlinearity on the LCU-prepared mix \cite{Khatri2024Quixer}.%

\subsection{Quantum Feed-Forward and Readout}
After LCU and QSVT we apply a second, independent PQC $U_{\mathrm{FF}}(\boldsymbol\phi)$ on the data register:
\begin{equation}
\ket{\psi} \;=\; U_{\mathrm{FF}}\, P_{\boldsymbol c}\!\big(M(\tilde{\boldsymbol b})\big)\, \ket{0^q}.
\label{eq:state}
\end{equation}
Unlike the embedding ansatz (shared per token), $U_{\mathrm{FF}}$ operates \emph{once per window} on the mixed state, mirroring the role of position-wise feed-forward in Transformers but executed in quantum state space. Decoupling representation construction through the embedding ansatz from representation consumption through the quantum feed-forward increases modeling capacity without increasing token-wise parameters. This decoupling naturally leads to a clear transition from quantum feature transformation to classical readout and classification.

For output, we perform XYZ multi-axis readout. For each qubit, we measure the expectation values w.r.t.\ $X$, $Y$, and $Z$, yielding a feature vector $\mathbf o\in\mathbb{R}^{3q}$. A two-layer MLP with dropout maps $\mathbf o$ to logits $\mathbf z\in\mathbb{R}^{C}$. This readout-and-head design finally adapts the head to classification.

\subsection{Objective, Regularization, and Optimization}
We minimize the cross-entropy $\mathrm{CE}(z,y)$ over logits $z$ and labels $y$, augmented with stability regularizers. Our objective is:
\begin{equation}
\mathcal{L} = \mathrm{CE}(z,y) + \mathrm{PSR} + \mathrm{L1C}.
\label{eq:loss}
\end{equation}

The post-selection regularizer $\mathrm{PSR}$ steers the average success probability toward a target $\tau \in (0,1)$:
\begin{equation}
\mathrm{PSR} = \lambda_{\mathrm{ps}}\Big(\overline{\big\|P_{\mathbf c}(M)\lvert 0^q\rangle\big\|^2} - \tau\Big)^2.
\label{eq:psr}
\end{equation}

Here, $\overline{\|P_{\mathbf c}(M)\lvert 0^q\rangle\|^2}$ is the mean state norm (a proxy for post-selection success), $M$ is a parameterized quantum circuit acting on the $q$-qubit zero state, and $P_{\mathbf c}$ is the projector onto the measurement outcome $\mathbf c$. The overline denotes an average (e.g., over minibatch examples and/or circuit shots), and the regularizer $\mathrm{PSR}$ nudges this mean success probability toward a target $\tau \in (0,1)$.

The $\ell_1$ constraint $\mathrm{L1C}$ softly encourages a unit-sum mixing vector $\tilde{\mathbf b}$:
\begin{equation}
\mathrm{L1C} = \lambda_{\ell_1}\big(\|\tilde{\mathbf b}\|_{1} - 1\big)^2.
\label{eq:l1c}
\end{equation}

In practice we enforce $\ell_1$ normalization in the forward pass (thus $\lambda_{\ell_1}=0$) and use the mean state norm primarily for logging or light regularization. Unless otherwise specified, we train with AdamW and a cosine-annealed learning rate, following recent small-model quantum--classical NLP evaluation protocols.

\subsection{Resource Model and Complexity}
With window length $n$, QuixerClassifier uses $q{=}8$ data qubits, $\lceil \log_2 n\rceil$ control qubits, and a constant number of ancillae for QSVT and parity combination; total qubits scale as $\mathcal O(q+\log n)$. Let the embedding PQC have depth $\ell$ and the QSVT polynomial have degree $d$. Using standard controlled-gate constructions (or the ancilla-optimized variant), the gate complexity scales as
\begin{equation}
\mathcal O\!\big(d\,n\,q\,\ell\big),
\end{equation}
matching Quixer’s analysis and justifying our shallow-circuit, small-$q$ design.%

Compared with adaptive hybrids like AQCF that dynamically route between classical and quantum paths, our design fixes a compact quantum mixer and optimizes it end-to-end under the same small-model budget used for fair comparison on SST-2 and IMDB. The learnable LCU-QSVT mixer serves as an attention-like token mixer whose complex coefficients and polynomial parameters are directly tuned from data, yielding strong classification performance with only $8$ qubits.

\begin{table*}[t]
\centering
\small
\begin{tabular}{l c cccc|cccc}
\toprule
&  & \multicolumn{4}{c}{\textbf{SST-2 Dataset}} & \multicolumn{4}{c}{\textbf{IMDB Dataset}} \\
\cmidrule(lr){3-6} \cmidrule(lr){7-10}
\textbf{Method} & \textbf{Qubits} & \textbf{Acc.} & \textbf{Precision} & \textbf{Recall} & \textbf{F1-Score} & \textbf{Acc.} & \textbf{Precision} & \textbf{Recall} & \textbf{F1-Score} \\
& & (\%) & & & & (\%) & & & \\
\midrule
\multicolumn{10}{l}{\textit{Quantum Methods}} \\
DisCoCirC  & 20 & 53.44 & 0.536 & 0.536 & 0.534 & 50.00 & 0.750 & 0.500 & 0.333 \\
VQC Classifier & 20 & 52.29 & 0.262 & 0.500 & 0.343 & 54.36 & 0.544 & 0.544 & 0.543 \\
\midrule
\multicolumn{10}{l}{\textit{Classical Baseline}} \\
Classical Transformer & -- & 79.36 & 0.800 & 0.797 & 0.793 & 82.18 & 0.822 & 0.822 & 0.822 \\
\midrule
\multicolumn{10}{l}{\textit{Hybrid Methods}} \\
Hybrid Quantum CNN & 20 & 72.02 & 0.729 & 0.715 & 0.714 & 77.91 & 0.781 & 0.779 & 0.779 \\
Quantum Transformer & 20 & 79.59 & 0.796 & 0.797 & 0.796 & 82.30 & 0.829 & 0.823 & 0.822 \\
AQCF & 20 & 81.88 & 0.819 & 0.818 & 0.818 & 86.30 & 0.863 & 0.863 & 0.863 \\
\midrule
\multicolumn{10}{l}{\textit{Proposed Method}} \\
CLAQS (Ours) & 8 & \textbf{91.64} & \textbf{0.915} & \textbf{0.915} & \textbf{0.915} & \textbf{87.08} & \textbf{0.871} & \textbf{0.871} & \textbf{0.871} \\
\bottomrule
\end{tabular}
\caption{Performance comparison on SST-2 and IMDB sentiment analysis benchmarks. “--” indicates that the reported qubit count is unspecified.}
\label{tab:main_results}
\end{table*}

\section{Experiments and Results}
\subsection{Tasks and datasets}
We evaluate CLAQS on two standard sentiment classification benchmarks: SST-2, short sequences from the Stanford Sentiment Treebank \cite{socher2013recursive} and IMDB, long reviews from \citet{maas2011learning}. We use the official train/validation/test splits and report Accuracy, Precision, Recall, and macro F1 on the test sets. These benchmarks are widely used to assess token-mixing mechanisms and attention alternatives in NLP. They further allow us to probe the sequence-length axis most relevant to token mixing.

\begin{tcolorbox}[
    colback=blue!3!white,
    colframe=black,
    boxrule=0.8pt,
    arc=3pt,
    left=6pt,
    right=6pt,
    top=4pt,
    bottom=4pt,
    title=\textbf{RQ1. Model Effectiveness},
    coltitle=white,
    fonttitle=\bfseries,
    breakable
]
Does the proposed \textbf{CLAQS} architecture outperform classical and hybrid baselines in text classification, 
given an equal or smaller parameter and qubit budget?  
\end{tcolorbox}

\vspace{4pt}

\begin{tcolorbox}[
    colback=blue!3!white,
    colframe=black,
    boxrule=0.8pt,
    arc=3pt,
    left=6pt,
    right=6pt,
    top=4pt,
    bottom=4pt,
    title=\textbf{RQ2. Component Contribution},
    coltitle=white,
    fonttitle=\bfseries,
    breakable
]
Which components of the \textbf{CLAQS} mixer (LCU, QSVT, and quantum feed-forward) 
most contribute to the observed performance gains, and how do aggregation or regularization schemes influence these results?  
\end{tcolorbox}

\subsection{Implementation Details}
We implement all quantum components in \textbf{TorchQuantum}, explicitly injecting rotation angles and realizing \textbf{LCU} and \textbf{QSVT} as dedicated differentiable layers. 
For classical simulation, we follow \textbf{Quixer}'s efficient strategy: repeatedly evaluating $M$ and applying $P_{\boldsymbol{c}}$ through compositional updates rather than materializing full block-encoding circuits. 
Parameter initialization employs \textbf{Xavier} for $W_E$, small uniform noise for PQC angles, and near-isotropic initialization for $\boldsymbol{b}$ and $\boldsymbol{c}$, followed by end-to-end optimization. 
During the forward pass, $\boldsymbol{b}$ is batch-broadcast and \textbf{L$^1$-normalized} to prevent coefficient explosion or collapse, ensuring stable success probabilities. 

All models share the same tokenization and official train/validation/test splits as their respective datasets. 
The \textbf{CLAQS} mixer (LCU + QSVT) is optimized jointly with the two-stage PQC and the lightweight MLP head. 
We adopt the small-model evaluation protocol commonly used in recent quantum--classical NLP research, and conduct all experiments on an NVIDIA RTX~A5000 (24\,GB) GPU.

\begin{table*}[t]
\centering
\small
\begin{tabular}{lcccc|cccc}
\toprule
\multirow{2}{*}{\textbf{Model / Variant}} & \multicolumn{4}{c}{\textbf{SST-2}} & \multicolumn{4}{c}{\textbf{IMDB}} \\
\cmidrule(lr){2-5} \cmidrule(lr){6-9}
 & Acc (\%) & Prec. & Rec. & F1 & Acc (\%) & Prec. & Rec. & F1 \\
\midrule
\textbf{Full CLAQS (Ours)}        & \textbf{91.64} & 0.915 & 0.915 & 0.915 & \textbf{87.08} & 0.871 & 0.871 & 0.871 \\
\midrule
\multicolumn{9}{l}{\textit{Ablation Study 1: Aggregation Mechanism}} \\
mean\_logits                       & 78.90 & 0.790 & 0.789 & 0.789 & 84.39 & 0.844 & 0.844 & 0.844 \\
attention pooling                  & 79.82 & 0.798 & 0.798 & 0.798 & 83.63 & 0.836 & 0.836 & 0.836 \\
\midrule
\multicolumn{9}{l}{\textit{Ablation Study 2: Mask and Quantum Regularization}} \\
mean\_logits                       & 80.39 & 0.804 & 0.804 & 0.804 & 85.33 & 0.853 & 0.853 & 0.853 \\
+mask +LCU/QSVT regs               & 70.24 & 0.702 & 0.702 & 0.702 & 85.83 & 0.858 & 0.858 & 0.858 \\
\bottomrule
\end{tabular}
\caption{
\textbf{Ablation Results.}
Top rows report full CLAQS performance.
Middle block (\textit{Ablation Study 1}) compares aggregation mechanisms (mean\_logits vs. attention) under identical hyperparameters.
Bottom block (\textit{Ablation Study 2}) examines effects of measurement masking and quantum-side regularization (LCU L$^1$, QSVT smoothness L$^2$ diff, overall QSVT L$^2$ penalties).
}
\label{tab:ablation_combined}
\end{table*}

\begin{table*}[t]
\centering
\begin{tabular}{l l c c}
\toprule
\textbf{Method} & \textbf{Configuration} & \textbf{Attention Params} & \textbf{Qubits} \\
\midrule
Classic Transformer & $d_\text{model}{=}256$, heads{=}8 & 263{,}168 & -- \\
AQCF (default) & embed\_dim{=}128, $L{=}2$ & 8{,}268 & 20 \\
\textbf{Ours (IMDB)} & window{=}256, degree{=}5 & \textbf{454} & \textbf{8} \\
\textbf{Ours (SST-2)} & window{=}128, degree{=}5 & \textbf{326} & \textbf{8} \\
\bottomrule
\end{tabular}
\caption{
\textbf{Comparison of trainable attention parameters and qubit usage.}
For \textbf{Ours}, the parameter count excludes token embedding parameters and includes only 
LCU coefficients ($2n$ for window length $n$), QSVT polynomial coefficients ($d{+}1{=}6$ for degree{=}5),
and quantum feed-forward parameters. For \textbf{AQCF}, 8{,}268 already aggregates all trainable attention parameters under the stated defaults, where $L$ is the number of transformer layers.
}
\label{tab:attention_qubits}
\end{table*}

\subsection{Main results}
Table~\ref{tab:main_results} summarizes the results on SST-2 and IMDB.
On \textbf{SST-2}, \textbf{CLAQS} reaches \textbf{91.64\%} accuracy (F1 = 0.915), outperforming both the \textit{Classical Transformer} (79.36\%) and the strongest hybrid baseline \textit{AQCF} (81.88\%) by \textbf{+12.28} and \textbf{+9.76} points, respectively.
On \textbf{IMDB}, CLAQS achieves \textbf{87.08\%} accuracy, surpassing \textit{AQCF} (86.30\%, +0.78), \textit{Quantum Transformer} (82.30\%, +4.78), and the \textit{Classical Transformer} (82.18\%, +4.90).
Fully quantum methods underperform on both datasets, highlighting the need for principled quantum–classical integration. 
These gains are particularly notable given our strict \textbf{8-qubit} budget and shallow circuits, aligning with the resource-conscious design advocated in prior work~\cite{Khatri2024Quixer,pan2025bridging}. 

\subsection{Ablation studies}

\paragraph{Ablation 1: Aggregation mechanism}
We compare the CLAQS mixer (LCU+QSVT+U\textsubscript{FF}) against two purely classical window aggregators that keep all other hyperparameters fixed: \textit{mean\_logits} and \textit{attention pooling}.
On \textbf{SST-2}, CLAQS improves over \textit{mean\_logits} by \textbf{+12.74} points (91.64 vs.\ 78.90) and over \textit{attention pooling} by \textbf{+11.82} points (79.82).
On \textbf{IMDB}, CLAQS also outperforms both baselines by \textbf{+2.69} (vs.\ mean\_logits = 84.39) and \textbf{+3.45} (vs.\ attention = 83.63) points.
This indicates that the \emph{quantum} mixing induced by LCU (complex, normalized coefficients) plus QSVT (learnable polynomial nonlinearity) provides consistent benefits over classical pooling, with larger margins on the shorter sequence regime (SST-2).

\paragraph{Ablation 2: Measurement mask and quantum-side regularization}
We investigate two factors: (i) a measurement \emph{mask} on the XYZ readout, and (ii) quantum-side \emph{regularization} (LCU L$^1$, QSVT smoothness via L$^2$ on coefficient differences, and an overall L$^2$ penalty on the polynomial). 
On \textbf{SST-2}, the masked+regularized variant drops to 70.24\% (vs.\ 91.64\%), suggesting that strong constraints can \emph{underfit} small datasets where flexible token interactions are beneficial.
On \textbf{IMDB}, the same variant slightly exceeds the classical \textit{mean\_logits} baseline (85.83 vs.\ 85.33) but remains below full CLAQS (87.08).
Overall, aggressive smoothing and masking can curb expressivity on short inputs, while mild smoothing may be tolerable or modestly helpful on longer reviews; 
the full CLAQS remains the most effective across both settings.

\subsection{Parameter Budget}
Table \ref{tab:attention_qubits} compares trainable attention parameters excluding embeddings and qubit usage across models. A single 256-d, 8-head Transformer attention block contains 263{,}168 parameters. By contrast, our CLAQS attention uses only 454 parameters on IMDB and 326 on SST-2, which is, roughly $580\times$ and $807\times$ fewer than the Transformer block, respectively.
Compared with AQCF (\textbf{8{,}268} parameters), CLAQS is about $18\times$ (IMDB) to $25\times$ (SST-2) smaller.

Transformer attention parameters grow quadratically with width and linearly with the number of layers. In contrast, CLAQS is independent of $d_{\text{model}}$ and the number of heads; its attention budget scales linearly with the window length $n$ and the polynomial degree $d$, plus a small constant from the quantum feed-forward:
\[
P_{\text{attn}}^{\text{CLAQS}} = O(n) + O(d) + O(1).
\]
Empirically, halving the window from $n{=}256$ to $n{=}128$ reduces the parameter count by exactly $128$ (\(454 \rightarrow 326\)), matching the linear dependence on $n$.

\subsection{Discussion}
\paragraph{Where do the gains come from?}
The main improvements arise from treating the attention-like mixer as a learnable quantum operator: 
(i) complex LCU coefficients (with L$^1$ normalization) provide phase-sensitive weighted mixing, and 
(ii) QSVT supplies a learnable polynomial nonlinearity that composes token unitaries into higher-order interactions, before a \emph{window-level} quantum feed-forward refines the state.
This design follows the block-encoding and QSVT principles of Quixer but exposes all mixer parameters to gradient-based learning and adapts the head to classification. 

Compared to AQCF, which adaptively routes between classical and quantum paths, CLAQS fixes a compact quantum mixer and focuses the learning capacity on its complex coefficients and polynomial, yielding stronger results under the same small-model budget. 

\paragraph{Resource considerations.}
All results are obtained with $q{=}8$ data qubits; the control register scales with $\lceil\log_2 n\rceil$, and a small number of ancillae support QSVT and parity combination.
This adheres to the $\mathcal{O}(q+\log n)$ qubit count and $\mathcal{O}(d\,n\,q\,\ell)$ gate complexity characterized for LCU+QSVT mixers. 
We emphasize that our ablations keep the qubit budget fixed to isolate the contribution of the mixer itself.

With a strict 8-qubit budget and shallow depth, CLAQS delivers sizable gains over classical pooling and strong hybrid baselines on SST-2, and consistent improvements on IMDB.
These results support a practical view of quantum components in NLP: compact qubit-based mixers can act as effective, learnable alternatives to classical attention within modern pipelines. Moreover, a limited number of qubits can effectively represent the input information, as a single quantum state inherently encodes multiple interdependent variables. In essence, a small set of qubits can span an informational space~\cite{yosida2012functional, royden1988real} equivalent to that of classical representations within the same dimensional framework, highlighting the expressive efficiency of quantum encoding.

\section{Conclusion}
We introduced CLAQS, a Compact Learnable All-Quantum Token Mixer with Shared-ansatz, demonstrating that a compact quantum module can effectively serve as an attention-like token mixer for text classification. Architecturally, CLAQS exposes both the LCU mixing weights and the QSVT polynomial as learnable complex parameters. It separates representation construction from consumption through a two-stage PQC design with a shared ansatz-14 embedding stage followed by a window-level quantum feed-forward stage. Finally, it applies L$^1$ normalization on the LCU coefficients to stabilize training and prevent parameter divergence. With only $q{=}8$ data qubits and shallow circuits, CLAQS attains \textbf{91.64\%} accuracy on SST‑2 and \textbf{87.08\%} on IMDB, outperforming a classical Transformer and strong hybrid baselines (including AQCF), while remaining within realistic NISQ‑style resource budgets.

Our analyses support three takeaways. First, making the LCU and QSVT learnable is key since phase-aware LCU mixing combined with a polynomial nonlinearity consistently outperforms classical window aggregation such as mean or attention pooling. Second, the separation between the token‑shared embedding ansatz and the window‑level quantum feed‑forward improves adaptation to the task without increasing token‑wise parameters. Third, the gains persist under a strict 8‑qubit budget, indicating that small quantum mixers can be practical components in modern NLP pipelines.

\paragraph{Future directions.}
In future work, we will validate CLAQS on quantum hardware with shot-based readout under realistic noise. We will also broaden the empirical scope beyond sentiment analysis to natural language inference, topic classification, and token-level tasks, and explore scaling via multi-head and multi-layer variants alongside parameter-efficient training. In parallel, we will examine constrained and Chebyshev-based QSVT polynomials to understand their effects on optimization stability and calibration. Finally, we aim to develop a theoretical account of post-selection success and gradient scaling in compact QSVT–LCU mixers. More broadly, we view CLAQS as a step toward quantum–classical co-design, where compact qubit mixers can serve as learnable alternatives to attention under tight resource constraints.

\bibliography{main.bib}

\begin{thebibliography}{22}
\providecommand{\natexlab}[1]{#1}

\bibitem[{Alexander and Widdows(2022)}]{alexander2022quantum}
Aaranya Alexander and Dominic Widdows. 2022.
\newblock Quantum text encoding for classification tasks.
\newblock In \emph{2022 IEEE/ACM 7th Symposium on Edge Computing (SEC)}, pages 355--361. IEEE.

\bibitem[{Chen et~al.(2025)Chen, Zhao, Feng, Chen, Lin, and Lin}]{chen2025quantum}
Fu~Chen, Qinglin Zhao, Li~Feng, Chuangtao Chen, Yangbin Lin, and Jianhong Lin. 2025.
\newblock Quantum mixed-state self-attention network.
\newblock \emph{Neural Networks}, 185:107123.

\bibitem[{Cherrat et~al.(2024)Cherrat, Kerenidis, Mathur, Landman, Strahm, and Li}]{cherrat2024quantum}
El~Amine Cherrat, Iordanis Kerenidis, Natansh Mathur, Jonas Landman, Martin Strahm, and Yun~Yvonna Li. 2024.
\newblock Quantum vision transformers.
\newblock \emph{Quantum}, 8(arXiv: 2209.08167):1265.

\bibitem[{Coecke(2021)}]{coecke2021mathematics}
Bob Coecke. 2021.
\newblock The mathematics of text structure.
\newblock In \emph{Joachim Lambek: The Interplay of Mathematics, Logic, and Linguistics}, pages 181--217. Springer.

\bibitem[{Gruzdeva et~al.(2025)Gruzdeva, Iurev, Bessmertny, Khrennikov, and Alodjants}]{gruzdeva2025quantum}
Anastasia~S Gruzdeva, Rodion~N Iurev, Igor~A Bessmertny, Andrei~Y Khrennikov, and Alexander~P Alodjants. 2025.
\newblock A quantum-like approach to semantic text classification.
\newblock \emph{Entropy}, 27(7):767.

\bibitem[{Gu and Dao(2023)}]{gu2023mamba}
Albert Gu and Tri Dao. 2023.
\newblock Mamba: Linear-time sequence modeling with selective state spaces.
\newblock \emph{arXiv preprint arXiv:2312.00752}.

\bibitem[{Khatri et~al.(2024)Khatri, Matos, Coopmans, and Clark}]{Khatri2024Quixer}
Nikhil Khatri, Gabriel Matos, Luuk Coopmans, and Stephen Clark. 2024.
\newblock Quixer: A quantum transformer model.
\newblock \emph{arXiv preprint arXiv:2406.04305}.

\bibitem[{Lee-Thorp et~al.(2022)Lee-Thorp, Ainslie, Eckstein, and Ontanon}]{lee-thorp-etal-2022-fnet}
James Lee-Thorp, Joshua Ainslie, Ilya Eckstein, and Santiago Ontanon. 2022.
\newblock \href {https://doi.org/10.18653/v1/2022.naacl-main.319} {{FN}et: Mixing tokens with {F}ourier transforms}.
\newblock In \emph{Proceedings of the 2022 Conference of the North American Chapter of the Association for Computational Linguistics: Human Language Technologies}, pages 4296--4313, Seattle, United States. Association for Computational Linguistics.

\bibitem[{Li et~al.(2024)Li, Zhao, and Wang}]{li2024quantum}
Guangxi Li, Xuanqiang Zhao, and Xin Wang. 2024.
\newblock Quantum self-attention neural networks for text classification.
\newblock \emph{Science China Information Sciences}, 67(4):142501.

\bibitem[{Lorenz et~al.(2023)Lorenz, Pearson, Meichanetzidis, Kartsaklis, and Coecke}]{lorenz2023qnlp}
Robin Lorenz, Anna Pearson, Konstantinos Meichanetzidis, Dimitri Kartsaklis, and Bob Coecke. 2023.
\newblock Qnlp in practice: Running compositional models of meaning on a quantum computer.
\newblock \emph{Journal of Artificial Intelligence Research}, 76:1305--1342.

\bibitem[{Maas et~al.(2011)Maas, Daly, Pham, Huang, Ng, and Potts}]{maas2011learning}
Andrew Maas, Raymond~E Daly, Peter~T Pham, Dan Huang, Andrew~Y Ng, and Christopher Potts. 2011.
\newblock Learning word vectors for sentiment analysis.
\newblock In \emph{Proceedings of the 49th annual meeting of the association for computational linguistics: Human language technologies}, pages 142--150.

\bibitem[{Meichanetzidis et~al.(2023)Meichanetzidis, Toumi, de~Felice, and Coecke}]{meichanetzidis2023grammar}
Konstantinos Meichanetzidis, Alexis Toumi, Giovanni de~Felice, and Bob Coecke. 2023.
\newblock Grammar-aware sentence classification on quantum computers.
\newblock \emph{Quantum Machine Intelligence}, 5(1):10.

\bibitem[{Pan et~al.(2025)Pan, Jiang, Chen, Li, Zhao, Zhao, Abate, Wang, and Liu}]{pan2025bridging}
Yi~Pan, Hanqi Jiang, Junhao Chen, Yiwei Li, Huaqin Zhao, Lin Zhao, Yohannes Abate, Yingfeng Wang, and Tianming Liu. 2025.
\newblock Bridging classical and quantum computing for next-generation language models.
\newblock \emph{arXiv preprint arXiv:2508.07026}.

\bibitem[{Payares et~al.(2023)Payares, Puertas, and Martinez-Santos}]{payares2023quantum}
Esteban Payares, Edwin Puertas, and Juan~C Martinez-Santos. 2023.
\newblock Quantum n-gram language models for tweet classification.
\newblock In \emph{2023 IEEE 5th International Conference on Cognitive Machine Intelligence (CogMI)}, pages 69--74. IEEE.

\bibitem[{Poli et~al.(2023)Poli, Massaroli, Nguyen, Fu, Dao, Baccus, Bengio, Ermon, and R{\'e}}]{poli2023hyena}
Michael Poli, Stefano Massaroli, Eric Nguyen, Daniel~Y Fu, Tri Dao, Stephen Baccus, Yoshua Bengio, Stefano Ermon, and Christopher R{\'e}. 2023.
\newblock Hyena hierarchy: Towards larger convolutional language models.
\newblock In \emph{International Conference on Machine Learning}, pages 28043--28078. PMLR.

\bibitem[{Royden and Fitzpatrick(1988)}]{royden1988real}
Halsey~Lawrence Royden and Patrick Fitzpatrick. 1988.
\newblock \emph{Real analysis}, volume~32.
\newblock Macmillan New York.

\bibitem[{Smaldone et~al.(2025)Smaldone, Shee, Kyro, Farag, Chandani, Kyoseva, and Batista}]{smaldone2025hybrid}
Anthony~M Smaldone, Yu~Shee, Gregory~W Kyro, Marwa~H Farag, Zohim Chandani, Elica Kyoseva, and Victor~S Batista. 2025.
\newblock A hybrid transformer architecture with a quantized self-attention mechanism applied to molecular generation.
\newblock \emph{Journal of Chemical Theory and Computation}, 21(10):5143--5154.

\bibitem[{Socher et~al.(2013)Socher, Perelygin, Wu, Chuang, Manning, Ng, and Potts}]{socher2013recursive}
Richard Socher, Alex Perelygin, Jean Wu, Jason Chuang, Christopher~D Manning, Andrew~Y Ng, and Christopher Potts. 2013.
\newblock Recursive deep models for semantic compositionality over a sentiment treebank.
\newblock In \emph{Proceedings of the 2013 conference on empirical methods in natural language processing}, pages 1631--1642.

\bibitem[{Vaswani et~al.(2017)Vaswani, Shazeer, Parmar, Uszkoreit, Jones, Gomez, Kaiser, and Polosukhin}]{vaswani2017attention}
Ashish Vaswani, Noam Shazeer, Niki Parmar, Jakob Uszkoreit, Llion Jones, Aidan~N Gomez, {\L}ukasz Kaiser, and Illia Polosukhin. 2017.
\newblock Attention is all you need.
\newblock \emph{Advances in neural information processing systems}, 30.

\bibitem[{Yang et~al.(2022)Yang, Qi, Chen, Tsao, and Chen}]{yang2022bert}
Chao-Han~Huck Yang, Jun Qi, Samuel Yen-Chi Chen, Yu~Tsao, and Pin-Yu Chen. 2022.
\newblock When bert meets quantum temporal convolution learning for text classification in heterogeneous computing.
\newblock In \emph{ICASSP 2022-2022 IEEE International Conference on Acoustics, Speech and Signal Processing (ICASSP)}, pages 8602--8606. IEEE.

\bibitem[{Yosida(2012)}]{yosida2012functional}
K{\^o}saku Yosida. 2012.
\newblock \emph{Functional analysis}, volume 123.
\newblock Springer Science \& Business Media.

\bibitem[{Zhao et~al.(2024)Zhao, Shi, and Li}]{zhao2024qksan}
Ren-Xin Zhao, Jinjing Shi, and Xuelong Li. 2024.
\newblock Qksan: A quantum kernel self-attention network.
\newblock \emph{IEEE Transactions on Pattern Analysis and Machine Intelligence}.

\end{thebibliography}


\end{document}